\documentclass[sigconf,10pt]{acmart}
\usepackage{graphicx}	
\usepackage{amssymb,blkarray}
\usepackage{amsmath}
\usepackage{xcolor}
\usepackage{balance}
\usepackage{amsthm}
\renewcommand\footnotetextcopyrightpermission[1]{} % removes footnote with conference information in first column
\newcommand{\ie}{{\em i.e., }}
\newcommand{\eg}{{\em e.g., }}
\newcommand{\etc}{{\em etc }}
\newlength{\figwidth}
\usepackage{algcompatible,newfloat}
\newtheorem{theorem}{Theorem}
\newtheorem{definition}{Definition}
\DeclareFloatingEnvironment[
    fileext=loa,
    listname=List of Algorithms,
    name=Algorithm,
]{new_algorithm}

\newcommand{\ER}{Erd{\H{o}}s-R{\'e}nyi }

\title{Bayesian inference of network structure from information cascades}
\author{Caitlin Gray}
\email{caitlin.gray@adelaide.edu.au}
\affiliation{School of Mathematical Sciences, University of Adelaide}
\affiliation{ARC Centre of Excellence for Mathematical \& Statistical  Frontiers}
\author{Lewis Mitchell}
\affiliation{School of Mathematical Sciences, University of Adelaide}
\affiliation{ARC Centre of Excellence for Mathematical \& Statistical  Frontiers}
\author{Matthew Roughan}
\affiliation{School of Mathematical Sciences, University of Adelaide}
\affiliation{ARC Centre of Excellence for Mathematical \& Statistical  Frontiers}

\begin{abstract}
Contagion processes are strongly linked to the network structures on which they propagate, and learning these structures is essential for understanding and intervention on complex network processes such as epidemics and (mis)information propagation. However, using contagion data to infer network structure is a challenging inverse problem. In particular, it is imperative to have appropriate measures of uncertainty in network structure estimates, however these are largely ignored in most machine-learning approaches. We present a probabilistic framework that uses samples from the distribution of networks that are compatible with the dynamics observed to produce network and uncertainty estimates. We demonstrate the method using the well known independent cascade model to sample from the distribution of networks $P(G)$ conditioned on the observation of a set of infections $C$. We evaluate the accuracy of the method by using the marginal probabilities of each edge in the distribution, and show the benefits of quantifying uncertainty to improve estimates and understanding, particularly with small amounts of data.

\end{abstract}
\begin{document}
\maketitle

\section{Introduction}

%INTRO SENTENCE ABOUT NETWORK STRUCTURE.  
Contagion phenomena, whether information flow, cascading power grid failures or disease epidemics, are inherently linked to the network structure on which they propagate.  Often, we observe dynamics without knowledge of the underlying network structure; either because it is unobservable or because the dynamics are easier or cheaper to record than the underlying network.  For example, it may be difficult to observe social structure as connections, like friendships, may are invisible or ill-defined. However, dynamics on these networks often are observable; for example, people posting the same `meme', buying the same product or sharing the same link.  This spatio-temporal information is inherently linked to the underlying network structure and can be used to reconstruct and predict paths of information flow.  Reconstruction of the network from these observations can help to understand how network structure interacts with dynamics and to predict or control the information flow.  

%
%However, many of the dynamics observed on networks, particularly those involving human behaviour, are inherently uncertain.  There is uncertainty in the cascade dynamics that are observed, \eg missing data, incorrect time recordings, as well as the underlying network structure. 

Here, we address the problem of inferring network structure from observations of information diffusion, accounting for the inherent uncertainties that arise in inference and data observations.  To achieve this we utilise Bayesian methods that quantify uncertainty to better understand and utilise network predictions.

There are many sources of uncertainty in social networks.  These networks are often dynamic and difficult to observe, and there has been significant effort to improve network estimates from noisy and missing data \cite{Farine-15, Koskinen-13,  Kossinets-06, Newman-18, Butts-03}. %Networks change often, can be difficult to define and observation processes often result in noisy estimates that obscure the true underlying network.
The use of an indirect observation, such as dynamics on the network, increase the uncertainty in the network estimates. When we can only achieve an estimate of the network through the lens of the cascade processes it is important to consider the uncertainty in the cascade dynamics, \eg missing data, incorrect time recordings, as well as heterogeneity in observability.

Probabilistic techniques are ideal to solve problems and generate insights in this context as they explicitly include such uncertainties. In this paper, we develop a probabilistic approach to the network inference problem of learning the underlying network structure from information cascades. 

We begin with the well-known independent cascade model of information propagation \cite{Kempe-03} and develop a Bayesian inference method based on Markov Chain Monte Carlo (MCMC) to infer links between individuals. We seek a distribution of graphs that can facilitate the observed cascades to not only find good estimates of the underlying graph, but also use the knowledge of the underlying distribution to facilitate informed decision making and promote further understanding of the uncertainties in these processes.

The method samples from $P(G|C)$: the distribution of networks $P(G)$ conditioned on the observation of a set of infections $C$. Point estimates and associated uncertainty about the existence of edges can then be obtained. By quantifying the uncertainty, estimates can be obtained with small amounts of data, and known information about edges can be easily incorporated. 

We demonstrate the method using the independent cascade model; however, it can easily be extended to any cascade model where the likelihood of a cascade can be evaluated for a given network. We show that Bayesian techniques can produce estimates when simulating only limited data where other methods, such as NETINF \cite{Gomez-12} can not produce a result.  The method outperforms the baseline for synthetic and real world networks. 

The main contributions of this paper are the development of a probabilistic technique to sample from the distribution of underlying graphs in order to recover network structure and to provide uncertainty quantification that can be used to inform decisions and understand the interaction between spreading process and network structures.

\section{Background and related work}
The network inference problem is an active area of research that aims to infer the links or transmission probabilities using both model and model-free inference.  Only recently have the uncertainties in these types of estimates begun to be quantified.

There is a significant body of work on inferring information about the underlying network structure from observations of information diffusion.  These methods are largely based in a maximum likelihood estimation (MLE) or expectation maximisation (EM) frameworks with a variety of optimisation strategies employed to find a good estimate of the network structure.
%Broadly these fall into two categories: inference of the network structure itself and inference of information about the network, \eg link strengths, node influence and diffusion mechanisms/pathways. 

A seminal work on network inference is the well known NETINF algorithm that uses submodular optimisation to infer the underlying graph structure.  Numerous extensions incorporate prior information about the underlying graph structure such as sparsity, motif frequency, community structure, \etc \cite{He-17, Ramezani-17, MADNI-18}.  Many of these are ad-hoc extensions to include prior information.  Additionally, algorithms have been developed to infer the strength of connections in a network \cite{NETRATE, Braunstein-19},  understanding the heterogeneous influence of edges on different topics \cite{He-17, KernelCascade, Wang-14} and derive bounds on the reliability of the methods \cite{Netrapalli-12}.

We require a general probabilistic framework for inference that incorporates uncertainty. Bayesian methods can provide this framework to infer distributions and quantify uncertainty, as well as include prior information about the network structure and cascade process.
 
Bayesian inference of networks from observation of network dynamics emerged in the the stochastic epidemic literature to recover parameters of underlying network models, \eg  the underlying parameter $p$ of a Bernoullli random graph, in addition to the epidemiological model parameters  \cite{Britton-02, Dutta-18}.  More recently, the benefits of Bayesian techniques to quantify uncertainty in networks, often due to collection methods, have been identified \cite{Newman-18, Peixoto-18a}.  This extends naturally to the use of these methods in the estimation of the network structure and its properties from the observation of dynamics on the network.

Embar \emph{et~al.} propose a Bayesian framework for estimating properties of the network, \eg edge strengths as well as cascade properties, \eg propagation trees \cite{Embar-14}.  Other Bayesian approaches of network inference aim to infer the underlying network when the exact infection time is unknown in both standard and online algorithms \cite{Linderman-14, Shaghaghian-16, Shaghaghian-17}. Additionally, some approaches based on MLE use Bayesian techniques at intermediate steps to improve results \cite{vine-16}.  Only recently have Bayesian techniques have been introduced to quantify the uncertainty of network structure inference from observation of dynamics \cite{Crawford-15, DYFERENCE, Peixoto-19,Vajdi-18}. Ghalebi  \emph{et~al.} highlight the need for general probabilistic frameworks for inference problems and propose the algorithm DYFERENCE that samples edge and node probabilities in an online algorithm for dynamic network inference \cite{DYFERENCE}.  Peixoto \cite{Peixoto-19} recently proposed the most closely related work in an algorithm designed to jointly reconstruct network structure and community labels assuming a stochastic block model structure and highlights the benefits of recovering the full posterior distribution of networks.

It is worth noting that Bayesian methods are extensively used in sampling exponential random graph models (ERGMs) and fitting coefficients of motifs to observed networks \cite{Friel-11, Butts-18,Lusher-12}.  Fitting ERGMs can be extended to the network inference problem by observing information cascades instead of observing an existing network to infer ERGM coefficients.

\section{Graph MCMC for the network inference problem}
The network inference problem aims to learn the structure of an underlying network from the observation of transmissions over the network.  A set of these transmissions is a cascade and could be an information cascade, where individuals are transmitting information, or an infection cascade where individuals are contracting disease from their contact networks.  On a given underlying graph $G = (N,E)$ with node set $N$ and edge set $E$, a \emph{cascade} is a series of time ordered events where a node $i \in N$ becomes ``activated", by one of its neighbours $j$, where $(i,j) \in E$.   We observe a sequence, $n_1, \dots, n_k$ of activated nodes and their time of activation $t_1, \dots, t_k$.  However, we do not observe the parent of the transmission; that is, we can not observe who infected node $i$, only the time they were infected.

This leads to the following problem. Given a set of cascades $C={c_1, c_2, ..., c_k}$, where $c_i = {\{t_1,t_2,t_3,...\}}_i$ is the infection times of the nodes in cascade $i$, what is the underlying graph $G$ on which the process was observed?

Here we are interested in the distribution over all graphs that could have produced the observed cascades $C$.  That is, we are interested in recovering $P(G|C)$.  Directly recovering this distribution is infeasible for graphs with more than a handful of nodes as the dimension of the distribution increases exponentially.  Instead, we draw samples from this distribution using Bayesian methods and use these to recover the most probable network structures and a measure of their probability.

\subsection{Transmission Model}
We use the independent cascade (IC) model of information diffusion on the edges of $G$, analogous to the SI model in epidemiology.   The IC model \cite{Kempe-03} assumes that every node $u$ independently infects its neighbour $v$ with some probability $\beta$ of success.  Originally proposed as a discrete time model, we use the continuous time extension \cite{Gomez-12}. Each node $u$ attempts infection of inactive neighbour $v$ after some time $\Delta_{u,v}$, so  $t_v= t_u + \Delta_{u,v}$.

We define $P(u,v|c) = P(t_v = t_u + \Delta_{u,v} | c, t_u)$ to be the probability density of the time node $v$ will be activated by node $u$ given the activation time $t_u$ and the cascade sequence.  We assume here that this depends only on the times of activations of the two nodes. However, this can be altered to include other characteristics like the degree of nodes, the importance of friendship $(u,v)$, properties of the node or the content of the message being spread.  There is evidence of both exponential and power-law distributed waiting times for contagion spread \cite{Malmgren-08,Barabasi-05, Mathews-18} and either can be incorporated in the proposed method. We demonstrate the method using the exponential distribution for the time between activations:
\begin{equation}
P(u,v|c) =  \exp(-\Delta_{u,v} / \alpha),
\end{equation}
for some parameter $\alpha$. For simplicity we use $\alpha=1$ for synthetic datasets.

The model assumes that only one node $u$ actually activates node $v$ despite the possibility of multiple active neighbours. Therefore, the resulting cascades are a tree. We choose the continuous time independent cascade model to demonstrate the MCMC method as it is simple, well studied and captures much of the dynamics of information flow.  However, the method described below can be extended to other diffusion models, such as epidemiological models, discrete time IC models or Hawkes process models \cite{Hawkes-71}, where the likelihood can be evaluated.

%
%In practice, when a node $u$ is activated we sample the time $\Delta_{u,v} \sim P(u,v|c)$ that $u$ will wait before attempting to activate neighbour $v$.  If successful the activation time of $v$ is given by $t_u = t_v + \Delta_{u,v}$.  

\subsection{Likelihood}
We require the likelihood of the set of cascades $C$ given some underlying graph $G$.  We use a similar derivation as in \cite{Gomez-12}.  First consider that cascades propagate via a tree so each node will have a single parent. Each possible tree is a disjoint outcome for the cascade on the graph, so,
\begin{equation}
\label{graph_probability}
P(c|G) = \sum_{T \in \mathcal{T}(G)} P(c,T |G),
\end{equation}
where $\mathcal{T}(G)$ is the set of all possible trees and $P(c,T |G)$ is the probability cascade $c$ travels through tree $T$ on graph $G$.

Consider the likelihood that the cascade $c$ in graph $G = (V,E)$ propagated in the tree $T=(V_T, E_T)$.  At each edge the cascade propagated with probability $\beta$ and stopped with probability $1-\beta$, giving:
\begin{eqnarray}
\nonumber
P(c,T|G) &=& P(T | G) P(c | T,G)\\
\nonumber
P(c,T|G) &=& \prod_{(u,v) \in E_T} \beta  \prod_{u \in V, (u,x) \in E \backslash E_T} 1-\beta  \prod_{(u,v) \in E_T} P(u,v|c)\\
& = & \beta^q (1-\beta)^r  \prod_{(u,v) \in E_T} P(u,v|c),
\label{prob_of_cascade_and_tree}
\end{eqnarray}
where $q$ is the number of edges over which the cascade propagated, $q = |E_T| = V_T -1$ , and $r$ is the  number of edges in the graph that the cascade did not pass through.  Mathematically, $r = \left( \sum_{u \in V_T} d_{out}(u) \right) - q $, where $d_{out}(u)$ is the out-degree of node $u$. 

Now consider the cascade over all possible trees on the graph $G$.  We substitute (\ref{prob_of_cascade_and_tree}) into (\ref{graph_probability}), to give
%We must determine all possible ways the cascade can spread in a tree across $G$.
%\begin{equation}
%	P(c|G) = \sum_{T \in \mathcal{T}_c(G)} P(c|T) P(T|G)
%\end{equation}
\begin{equation*}
P(c|G) = \sum_{T \in \mathcal{T}_c(G)} \beta^q (1-\beta)^r \prod_{(u,v) \in E_T}  P(u,v|c),
\end{equation*}
where $\mathcal{T}_c(G)$ is the set of all possible connected trees on the subgraph of $G$ induced by the nodes in $c$.  Note that in general we sum over all possible trees, but if the tree is inconsistent with the observed data then $P(c|T,G)$ is zero, and hence we can sum over the trees in the set $\mathcal{T}_c(G)$. %We assume that all trees are equally likely (apriori) so $P(T|G) = 1/|\mathcal{T}_c(G)|$ 
%\change{We note that the probability of a tree being possible on a given a graph is either $1$ if the tree is a subset of the graph, or $0$ if it is not.  Therefore the $P(T|G)$ term is $1$ for all terms in the sum.}

Each tree in $\mathcal{T}_c(G)$ contains the nodes activated in $c$, so $q$ and $r$ are independent of $T$, and so
\begin{equation}
\label{prob_of_cascade_on_graph}
P(c|G) = \beta^q (1-\beta)^r \sum_{T \in \mathcal{T}_c(G)}   \prod_{(u,v) \in E_T}  P(u,v|c).
\end{equation}
For a set of independent cascades $C$ occurring on $G$ we have
\begin{equation*}
P(C|G) = \prod_{c \in C} P(c|G).
\end{equation*}
Equation (\ref{prob_of_cascade_on_graph}) would naively require the sum over all possible spanning trees, which can be super-exponential in the size of $G$. However, as suggested in \cite{Gomez-12}, Kirchoff's matrix tree theorem, extended to directed trees by Tutte \cite{Tutte-48} allows this calculation in polynomial time in $N$ by exploiting the properties of the Laplacian matrix of the network.  We provide the theorem below.
\newpage
\begin{definition}[Laplacian of directed multigraphs]
\label{laplacian}
If $G(V,E)$ is a directed graph, we define the Laplacian $L$ as an $n \times n$ matrix with entries:
\begin{eqnarray*}
L(G) & = &  \left\{
       \begin{array}{ll}
         \sum_k w_{k,i}  & \mbox{ if $i =j$}, \\
         -w_{i,j}   & \mbox{ if $i \neq j$ \& $(i,j) \in E$}, \\
         0 & \mbox{otherwise}.
       \end{array} 
       \right.
\end{eqnarray*}
\end{definition}
%\newpage
\begin{theorem}[Tutte \cite{Tutte-48}]
\label{Tutte_laplacian}
If graph $G$ has Laplacian $L$ as defined in Definition (\ref{laplacian}), then the sum over the weighted trees of graph $G$ with root at node $j$ is
\begin{equation*}
\sum_{T\in L(G)} \prod_{(k,l) \in T} w_{k,l} = \mathrm{det~}(L(G)_r),
\end{equation*}
where $T$ is each directed spanning tree in $G$ and $L(G)_r$ is created by removing the $r$-th (root node) row and column from $L(G)$.
\end{theorem}

In our formulation we set $w_{i,j}$ to $P(i,j|c)$.  As the trees $T$ are directed acyclic graphs (DAGs), the determinant is upper triangular and so the adjacency matrix is the product of the diagonals \cite{Anton-10}. This simplifies calculation time and storage from $\mathcal{O}(N^2)$ to $\mathcal{O}(N)$ as we only work with the diagonal elements.

\subsection{Bayesian Inference}
Given a cascade set of cascades $C$ - a set of lists of activation times of labelled nodes in $V$ we infer a distribution of graphs conditioned on the occurrence of these cascades.  Mathematically, given the cascade set $C$ we sample from the distribution $P(G | C)$.

From (\ref{prob_of_cascade_on_graph}), we have the cascade probability from a graph $P(G|C)$.  From Bayes' rule we can get the distribution of interest:
\begin{equation}
P(G | C) = \frac{P(C|G)P(G)}{P(C)}.
\end{equation}

We use the basic Metropolis-Hastings MCMC algorithm to sample from the posterior.  At each step we propose a new graph $G'$ from the old graph $G$ using the proposal distribution $Q(G'|G)$.  The algorithm accepts the move step with the ratio between the cascade on the new graph to the old graph with probability:
\begin{eqnarray}
\label{alpha}
\nonumber
%\alpha &=& \mathrm{min}\left(1,\frac{P(C | G')}{P(C|G)}\frac{P(G')}{P(G)}\right)\\
\alpha&=&\mathrm{min}\left(1,\frac{Q(G|G')P(G')}{Q(G'|G)P(G)} \prod_{c\in C}\frac{P(c | G')}{P(c|G)}\right).\end{eqnarray}
For a single cascade $c$, we use Theorem \ref{Tutte_laplacian} and (\ref{prob_of_cascade_on_graph}) to get

\begin{equation*}
\label{prob_of_cascade_ratio}
\frac{P(c|G')}{P(c|G)} = \beta^{q'-q} (1-\beta)^{r'-r}\frac{\mathrm{det} (L(G')_{n_1})}{\mathrm{det} (L(G)_{n_1})}.
\end{equation*}
%\change{Recall that,
%\begin{eqnarray*}
%q &=& |V_T| - 1,\\
%r &=& \sum_{u \in V_T} d_{out} (u) - q.
%\end{eqnarray*}}
Suppose the proposal changes one link $(i,j)$ from $G$ to $G'$.  The number of nodes in the spanning trees $|V_T|$ remains constant, so $q'=q$.  Assuming $t_i<t_j$, if node $i$ is in the cascade $d_{out}(i)$ increases when adding a link, and decreases when removing.
\begin{equation*}
R_c = r'-r
     = \left\{
       \begin{array}{ll}
          +1, & \mbox{ if $i$ in $c$ \& adding edge}, \\
          -1, & \mbox{ if $i$ in $c$ \& removing edge}, \\
          0,    & \mbox{ otherwise}.
       \end{array}
     \right.
\end{equation*}

%To determine $r$ we consider node degrees in the cascade.  When $i$ and $j$ are not in the cascade the degree of the nodes in the tree do not change, so, $r'-r = 0$. Assuming $t_i<t_j$ if $i$ is in the cascade $d_{out}(i)$ increases when adding a link, and decreases when removing. Summarising,

This term penalises the addition of edges in the cascade.  This not only promotes the desired sparse networks but also reduces the probability of edges that didn't propagate the cascade as edges are less likely.

A common proposal used when implementing MCMC on networks is to change a random node pair (i.e. creating a new edge or removing old edges) \cite{ Friel-11, Gray-19}. However, in sparse graphs non-edges are proposed much more often than edges, and the sampler wastes time proposing new edges that are likely to be rejected.  In order to improve mixing, the ``tie no-tie" (TNT) sampler is often used \cite{Friel-11,Hunter-08}.  The TNT sampler selects with equal probability the set of edges or set of non-edges, and then swaps a random node-pair in that set.  This improves convergence and aids mixing by proposing edge removals at a higher frequency, particularly when we begin our inference from a graph denser than our desired ensemble. Additionally, trying to remove edges often will explore different transmission pathways that rely on different links in the network.   Other proposals, for example edge flips and switches \cite{Greenhill-18} can also be used.

\begin{new_algorithm}
  \begin{algorithmic}[1]
  \REQUIRE $C = {c_1, c_2, \cdots c_{|C|}}$
  \STATE Generate $G^{(0)}$
  \FOR {$t=1...K$}
    \STATE Generate $G' \sim Q(G' | G^{(t-1)})$
	\STATE Take 
			$G^{(t)}  = \left\{
       \begin{array}{ll}
          G', & \mbox{ with probability } \alpha \\
          G^{(t-1)},    & \mbox{ with probabiltiy } 1-\alpha.
       \end{array}
     \right.$  \
	
where $\alpha = \text{min}\left(1,\frac{P(C|G')P(G)Q(\theta \vert \theta')}{P(C|G)P(G) Q(\theta' \vert \theta)}\right) $
    	\ENDFOR
    \caption{Bayesian inference of networks.}
    \label{MH_algorithm}
  \end{algorithmic}
\end{new_algorithm}
%Substituting into (\ref{alpha}) we get the acceptance probability:
%\begin{equation}
%\alpha = \mathrm{min}\left(1,\frac{P(G')}{P(G)}  \prod_{c\in C}
%(1-\beta)^{R_c}
%\frac{\mathrm{det} (L(G')_j)}{\mathrm{det} (L(G)_j)}\right)
%\end{equation}

For the subsequent analysis we will assume $\beta$ is known and show that the network recovery is not sensitive to this choice. Alternatively we could include $\beta$ in the inference by assuming a prior and determining the posterior estimate through a Gibbs' sampling step.  Unfortunately, the sparsity of the inferred network has dependence on both $\beta$ and $p$, and untangling these two parameters independently is difficult under this model.  These parameters could be estimated separately using other data, perhaps by sub-sampling a small section of the network or observing similar cascades in a known area of the network.  Nonetheless, we show that, for inference, results are not greatly affected by the chosen parameters.
\begin{figure*}
\centering
%\makebox[0pt]{\includegraphics[width = 1.2\textwidth]{/Users/caitlingray/Documents/InformationDiffusion/CodeProjects/NetworkInference3/graphMCMC_phoenix/plots/weightedadjacencyandROC_2.pdf}}
\makebox[0pt]{\includegraphics[width = 1.2\textwidth]{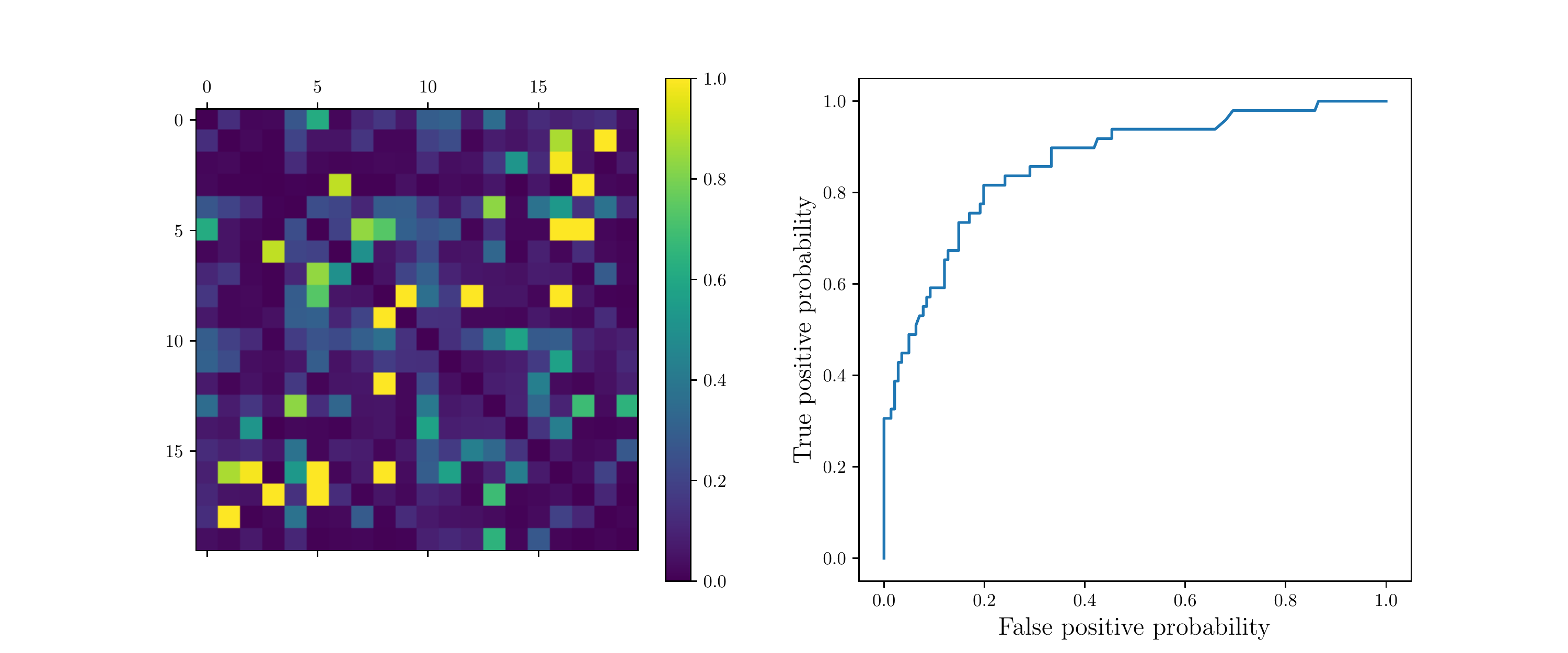}}
\caption{Left: Weighted matrix of estimated posterior edge probabilities.  Right: ROC curve of recovered edges from the posterior in the true network.}
\label{fig:recovered_network}
\end{figure*}
We place an \ER prior with parameter $p$ with independent edges on $G$, so $P(G')/P(G)$ is easily calculated. This prior serves as a regularisation term to promote sparse solutions. %Additionally, the uninformative prior allows us to investigate the link dependence that is introduced by the cascade observation and show that information sharing can be a mechanism for creating link dependent motifs, like edge pairs and triangles.

Overall the algorithm has two parts: generating the initial graph with associated probability and the iterative MCMC.  We generate the initial graph by connecting each time ordered pair in $C$. The complexity of the initial graph generation is $\mathcal{O}(|E_{init}|)$ where $|E_{init}|$ is the number of edges in the initial graph. The dominant part of the algorithm is the MCMC updates of graphs at each time step.  The method is $\mathcal{O}(K|C|)$, where $K$ is the number of iterations required for `convergence'.  Analysis of the graph density, average degree and clustering of graph samples along the MCMC chains with the TNT sampler suggest convergence scales like $K \sim N^2$.

\section{Experimental evaluation}
We test our algorithm on both synthetic to evaluate the effect the amount of data and the underlying network type has on performance, and extend this to inference of real networks. Next we describe the experimental setup. 

  %\change{For example, \ER newtorks show a linear increase in the average number of transmissions as the number of cascades increases. EITHER EXPAND OR DELETE}
It is known through work on disease and information transmissions that the structure of the underlying graph affects the cascades observed, even when using the same cascade model \cite{Gray-18, Watts-02}.  Therefore, we use a variety of undirected random graph models to represent the underlying graph $G$. We begin with the canonical \ER (ER) networks and then extend to more realistic networks with power law distributions using the Forest Fire model \cite{ForestFire}, core-periphery and hierarchical structures \cite{Clauset-08} using Kronecker networks \cite{Kronecker}.  %We work with sparse networks, as is often the case in real world situations. 
%In general, denser networks are more difficult to recover due to the increase in the number of possible transmission pathways, and so we often need more data to reliably infer denser networks.  
%\change{We show that if we keep the amount of data consistent (in terms of the fraction of edge transmissions we observe) the density of the network does not hinder recovery. (checking this now)}
%We demonstrate the performance on networks with $n=100$, and extend to testing different network types with $n=1024$. 

Cascades are simulated using the IC model with exponential waiting time ($\alpha = 1$). Intuitively, as the amount of data increases, \ie we observe more transmissions or cascades, the inference problem becomes easier.  We quantify the amount of data observed as the fraction, $f$, of the edges that are activated in the cascades,  \ie if all edges are activated at least once $f=1$.  This metric is inherently linked to the number of cascades observed, but ensures we do  over count cascades that do not provide new information. In most simulations we choose $\beta = 0.4$, to align with other works, including NETINF, and to ensure cascades are not too small, resulting in many cascades that have zero or one transmission, and not unrealistically large. For the purpose of this paper we assume that we have good estimates of $p$ and $\beta$, and show that even a poor estimate does not significantly impact results.  We select random seed nodes uniformly over the graph to ensure we have good coverage of the network even when the amount of cascade data is small.  

The proposed method provides samples from the distribution of $P(G|C)$, and so a key advantage of this method is the ability use these to quantify uncertainty and answer questions about the underlying distribution. Other methods provide a point estimate, $\hat{G}$, of the underlying graph. Despite the limitations of a simple point estimate such as this, obtaining an estimate from our posterior is useful, at least for comparison.  Due to the high dimensional posterior, we cannot guarantee that the maximum \emph{a posteriori} probability (MAP) estimate of the graph in our posterior is similar to the true maximum.  Of course, if we were specifically interested in this point estimate, then other methods using cleverly designed optimisation algorithms would be appropriate.  Nonetheless, we would like to be able to get some estimate of the graph and visualise the results in order to gain some understanding about the underlying graphs, and compare to other methods.  Therefore, to evaluate the solution quality we recover the marginal probabilities of the network edges, $q_{ij} = P((i,j) \in E | C)$. This can both provide important uncertainty quantification, and be used as a measure to determine how well this approximates the true network.

In general edges with high marginal probability are more likely to have been observed in the true graph $G^{*}$. We present these $q_{ij}$ in a weighted adjacency matrix (Figure \ref{fig:recovered_network}) and the most probable edges are used to create an estimate of the graph.  \footnote{If there is prior knowledge of the underlying network, such as the degree distribution, we could find the most probable edge set satisfying these constraints.}

The probabilistic nature of the inference provides the opportunity to use skill scores and probabilistic scoring techniques to compare the recovered distribution to the true network.  This is qualitatively different to most other approaches taken previously for this problem, which treat every problem as a `hard' classification of edge existence. Hence, in order to compare with other techniques we use the Receiver Operating Characteristic (ROC) curves that report the true positive rate and false positive rate of the method as the probability threshold varies. For example, Figure \ref{fig:recovered_network} shows the ROC curve for the posterior marginal probabilities for an undirected 20 node network.  We use two summaries of the ROC curves; the area under the curve (AUC) and the `false positive alarm' to demonstrate the recovery of the true edges in larger scenarios. 

The AUC provides an aggregate measure of classification performance across all thresholds, and is measured as the integral of the empirical ROC curve between $0$ and $1$. The false positive alarm measures the proportion of true edges correctly recovered before there are too many incorrect edges.  In many situations false positives are costly, either financially, \eg cost of investigating a false claim, or ethically, - \eg incorrect convictions.  Understanding the uncertainty in the estimate of the network, or probability of the edge, allows better decision making and application specific requirements on certainty. In this case we determine the true positive rate when 1\% of the edges recovered are false positives. 

%\change{To gain some understanding of the dependences in the network, we also look at the prevalence of edge pairs and triangles in the recovered distribution and compare this recovery to the underlying graph in the same way as above.}% WHY DO WE CARE ABOUT THESE????}%  Understanding the uncertainty in our estimates allows us to make better decisions and learn more about not only the network that exists but potential shadow pathways and links that play important roles in spreading diffusion.}% We investigate this by looking at highly probable links that are not in the underlying cascades and }

\subsection{Synthetic Data Results}
We begin by simulating cascades on synthetic networks generated from a variety of network models to investigate the impact of the amount of data and network type on performance.

%The data was simulated with $\beta = 0.4$ to ensure the cascades were large enough that we do not recover many edges easily from cascades with only one transmission. 

%\change{A benefit of the inference algorithm is the ability to provide insight into edges that are not observed in the cascades.  Edges that provide multiple pathways for the cascades to traverse are more likely, even when these edges are not directly observed.}   %in the cascade the algorithm shows this. Additionally, we are able to provide evidence for `non-edges'.  If node $i$ is activated, but node $j$ is not then this `non-transmission' reduces the likelihood of edge $ij$. 

%- ER networks (should be good)
%%- ER but with $p_{ij}$ instead (i.e. you know something in teh prior
%- Power law v Exponential waiting times of sharing
%- Forest Fire networks
%- kronecker networks

%The common testing networks are $n=1024$: not that big but still takes a while :/
%\begin{itemize}
%\item Kronecker graph	flat (ER model I think)
%\item core-periphery network [0.9,0.5;0.5,0.3]
%\item hierarchical community network [0.9,0.1;0.1;0.9]
%\item Forest Fire model [fwd 0.2,bwd 0.17]
%\end{itemize}
\subsubsection{Dependence on number of cascades}
It is intuitive that inferring the network becomes an easier problem when we have observed more data; however, in many scenarios data is limited.  Here we investigate the performance of the MCMC inference methods with varying amounts of data.
%Relationship between how well we recover the underlying network and the number/size of the cascades we observe

%The ability to recover the distribution of graphs that facilitate the cascades we observe is a major benefit of the algorithm.  
%In order to visualise the underlying graph distribution we investigate the edge probabilities in the recovered distribution.  Although we lose some information by assuming edge independence, it provides a basis to estimate of the underlying graph and compare to other methods.  Edge probabilities are also a natural way to visualise the uncertainty in estimation of the graph.

\begin{figure*}
\includegraphics[width=\textwidth]{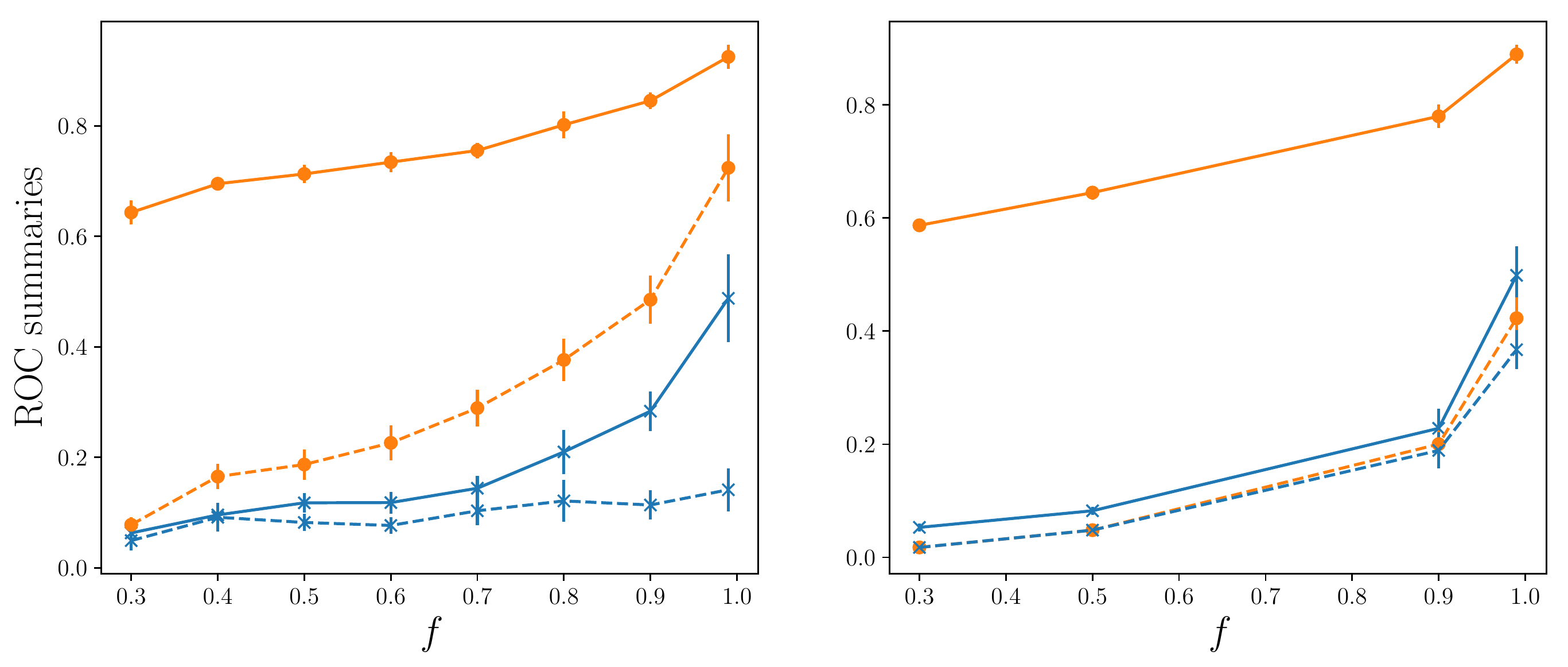}
\caption{ROC summary metrics for our method (solid lines) and NETINF (dashed) after observing cascades that cover a proportion $f$ of the edges. The orange curves (dots) show AUC and the blue (crosses) show false positive rate (blue crosses). Left: $n=100$. Right: $n=1000$}
\label{ROC_summaries_n100ER}
\end{figure*}

We begin with \ER graphs with $n=100$ and $n=1000$ with average degree $z=4$ .  The ROC curves from the edge probabilities extracted from the posterior samples are displayed in Figure \ref{ROC_summaries_n100ER}.  NETINF requires the number of edges $e$ in the resulting graph as an input. To create a comparative ROC curve, we report the true and false positive rates (TPR and FPR) for increasing $e$ and determine ROC summaries from this.
%curve (in the same way one makes a precision recall curve).
% (\change{results are very similar for other $z$ values that maintain a sparse graph})

Figure \ref{ROC_summaries_n100ER} shows the ROC summaries for the MCMC algorithm and NETINF for varying amounts of observed data. We observe that the MCMC algorithm performs well with relatively small amounts of data.  This is beneficial in many cases where data may be scarce.  When data is limited, the greedy algorithm used in NETINF does not return the requested number of edges and so performs poorly.  Understandably, most algorithms perform poorly under the false positive alarm for small amounts of data; however, when all edges are observed in at least one cascade ($f = 0.99$) the MCMC produces a good false alarm rate of over 40\%.

% \change{We demonstrate the benefits of using uncertainty quantification and prior knowledge when data is limited.}  

In practice, the underlying network is not known, and metrics like the false alarm rate cannot be observed. Therefore, it is difficult to choose how many edges are reliable in the recovered network. Using the MCMC method, unlike most other methods, we have quantified the uncertainty of each edge to inform such decisions when the underlying network is unknown.

%\subsubsection{Edge pairs \& Triangles}
%Social networks are known to have a high level of dependency between edges and transitivity (the prevalence of `triangles') is commonly observed.  In this method we use a prior that has assumes independent edges (\ER prior), so the dependence in the recovered distribution is a \change{property} of the cascades we observe.  To investigate this dependence we look at the edge pair and triangle posterior probabilities.
%
%\change{do this?} 

\subsubsection{Types of networks}
We test the performance of our inference scheme on different types of networks that display a variety of properties observed in real world networks.  
In this work we use an uninformative \ER prior that assumes links have an independent probability $p$ to control the sparsity of the solution. We show that despite this the algorithm can infer networks of varying types as the data informs the structure of the underlying graph.  We assume some prior knowledge of the edge density to choose $p$, but we show that the exact value has little effect on the network recovery.  In some cases we may have some prior knowledge of the underlying structure, perhaps the degree distribution, density or frequency of some motif, which can be incorporated in an informative prior to improve inference.

\begin{table}%[ht]
\centering
\caption{Inference results: $N=1000$} % title of Table
%\centering % used for centering table
\begin{tabular}{c c c} % centered columns (4 columns)

\hline\hline %inserts double horizontal lines
Network Type & Fraction & AUC \\ [0.5ex] % inserts table
%heading
\hline % inserts single horizontal line
\ER &  0.9 & 0.72 \\ % inserting body of the table
Forest Fire &  0.9 & 0.97\\ % inserting body of the table
Core-Periphery  & 0.9 & 0.76 \\ %\change{n=124}
Hierarchical  & 0.9 & 0.97 \\
%\change{Small World Network} &  \change{??} & \change{??} \\ % inserting body of the table
%Waxman/latent spatial model &  \change{??} & \change{??} \\ % inserting body of the table
\hline %inserts single line

\end{tabular}
\label{table:nonlin} % is used to refer this table in the text
\end{table}

There are considerable differences in the recoverability of networks of different types.  This is largely due to the nature of the cascades observed.  It is well known that the underlying structure impacts the nature of the cascades simulated using various cascade models \cite{Kempe-03,Gray-18}. Even for constant $\beta$ the underlying structure of the network has a dramatic impact on the size distribution of cascades we observe.  We simulate cascades until we observe a constant fraction $f=0.9$ of edges activated in each case. The average size and number of cascades required to achieve this changes between networks.  For \ER and core-periphery networks it is more likely that we observe fewer larger cascades covering many edges; compared to the Forest Fire and hierarchical networks in which large cascades are unlikely and so we observe many small cascades. Smaller cascades provide more precise information about edges as there are less possible paths the cascade could take, and so we see improved inference on networks that facilitate a smaller cascades sizes.

%\subsubsection{Impact of density and clustering of the network}
%\change{Perhaps include here a graph of impact of density and clustering on AUC results for smaller networks?}
%There is much work on fitting networks of different types to observed networks.  While this is hugely beneficial in some cases, it can give poor results if we choose a network type that is hugely different from the true network.  

\subsection{Sensitivity Analysis}
%\paragraph{What if we don't know the parameters $p$ and $beta$.}
There are two parameters used in the inference, the likelihood parameter $\beta$ and the prior probability $p$.  In the above analysis these parameters are assumed known, but in real implementations these will be estimates, $\hat{\beta}$ and $\hat{p}$ respectively. Therefore, it is useful to understand how sensitive our inference is on the estimates of these parameters.
%\begin{figure}
%\centering
%\includegraphics[width=0.6\columnwidth]{Figures/sensitivit_AUC_FP_p}
%\caption{The effects of incorrect parameters on the inference.  We see that incorrect $p$ values have very little effect on both AUC (blue crosses) and False positive alarm (purple squares).  The parameter $p$ (solid) is slightly more sensitive to large deviations from the true value; however, often we are interested in sparse graphs so $p$ is small. The true values of $p$ (black solid) and $\beta$ (black dashed) are shown.}
%\caption{The effects of incorrect parameters on the inference.  We see that incorrect $\beta$ (dotted) values have very little effect on both AUC (orange circles) and False positive alarm (purple squares).  The parameter $p$ (solid) is slightly more sensitive to large deviations from the true value; however, often we are interested in sparse graphs so $p$ is small. The true values of $p$ (black solid) and $\beta$ (black dashed) are shown.}
%\label{sensitivity}
%\end{figure}

\begin{figure}
\centering
\includegraphics[width=\columnwidth]{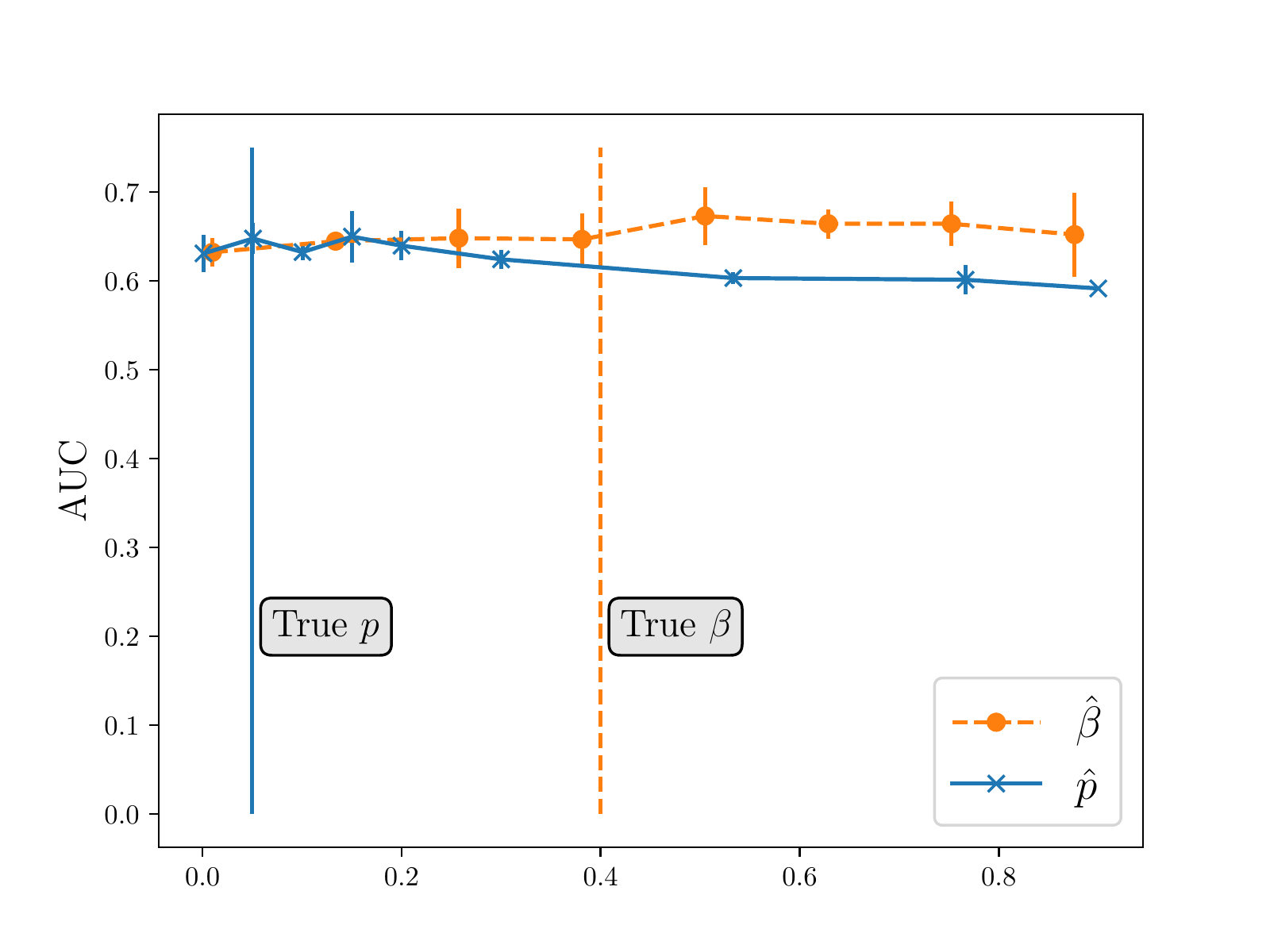}
\caption{The effects of incorrect parameters on the inference.  We see that incorrect $\hat{\beta}$ values (dashed orange) have very little effect on recovery in terms of AUC.  The parameter $\hat{p}$ (solid blue) is slightly more sensitive to large deviations from the true value. Vertical lines show the true $p$ (blue) and $\beta$ (orange).  Other ROC summaries such as false positive alarm (not shown) have the same trend.  }
\label{sensitivity}
\end{figure}

Recall that $\beta$ penalises the inclusion of links that cascades did not propagate through, while $p$ sets the prior probability of each edge and is used to promote sparse networks.  These two parameters are inherently linked to the number of edges in graphs of the posterior distribution.  However, in practice, these parameters do not significantly impact inference of the relative edge frequencies, and so do not impact the scale invariant AUC significantly. Figure \ref{sensitivity} shows the AUC as we vary the input $\hat{p}$ and $\hat{\beta}$ independently. Other ROC summaries demonstrate a similar trend.  There is a higher sensitivity to $\hat{p}$ as the denser graphs recovered at high $\hat{p}$ values allow multiple transmission pathways, many of which are not observed in the true underlying network. In general, the parameters alter density of the networks in the underlying distribution, but they do not significantly impact the estimated marginal probabilities of the edges.  

%\change{The edge density in the distributions are understandably impacted by the chosen values of $p$ and $\beta$. The distribution of density are shown for a variety of parameter values and give strong evidence for convergence of our algorithm to sampling from the distribution of interest.  High variance and asymmetric distributions are indicators the MCMC is still converging. }

\begin{table}%[ht]
\centering
\caption{Inference results on Email networks} % title of Table
%\centering % used for centering table
\begin{tabular}{c r r r r} % centered columns (4 columns)
\hline\hline %inserts double horizontal lines
Network Type & N & \begin{minipage}[t]{0.17\columnwidth}%
Undirected edges
\end{minipage} & \begin{minipage}[t]{0.19\columnwidth}%
Number of cascades
\end{minipage}  & AUC \\ [0.5ex] % inserts table
%heading
%\hspace{0.3cm}
\\[-1em]
\hline % inserts single horizontal line
Department 1 &  309 & 1938& 219 & 0.89\\ % inserting body of the table
Department 2 &  169 & 1045 & 143 & 0.93\\ % inserting body of the table
Department 3 & 89 & 973 & 34 & 0.74 \\
Department 4 &  142 &  833 & 129 & 0.95 \\ % inserting body of the table
\hline
Entire Network & 986 & 16064 & 234  &0.84\\
\hline %inserts single line
\end{tabular}
\label{table:realresults} % is used to refer this table in the text
\end{table}
\subsection{Experiments on real networks}

Next, we evaluate the proposed algorithm on a real world network of email communications \cite{SNAP} at a European Research Institute.  Email networks commonly observe cascades, for example spam emails, joke emails or event invitations.  We use the real underlying networks and simulate cascades, with $\beta = 0.2$, until 90\% of links are activated at least once, to be consistent with the previous experiment. This is significantly less information than used in some other works \cite{MADNI-18}.  We see that recovery in terms of AUC value can vary dramatically over the four departments. This is largely due to the different structures in each department that impact the properties of the cascades.  Department 3 has one giant component with a high density ($\sim0.2$), while the other departments are clustered with lower density ($< 0.1$).   Recovering denser networks is a harder inference problem as there are many more possible transmission pathways.  These results are significantly better than the NETINF results of $AUC \approx 0.5$ for all departments using the same data.  %Perhaps surprisingly, we recover the larger graph more effectively that some smaller departments in these experiments, due to the lower density of the total network compared to the individual departments. 

%\change{Compare this inference to other inference schemes (MADNI) which gives F1 = 0.6  $C=1000$ - what is an F1 score in our case?}

\section{Discussion}
Beyond improved performance, particularly when data is limited, there are many benefits and possible extensions to a Bayesian approach to this problem. We will discuss some of these below.

\paragraph{What effect does the prior knowledge give us?}
A major benefit of Bayesian inference is the ability to incorporate prior knowledge of all or part of the underlying distribution.  If some of the edges or edge dependencies in the underlying graph are known, perhaps from prior experiments or partial observations, this can be incorporated to provide a better estimate of the other edges.  This allows the above method to be used in missing link inference, when much of the network is known. Alternatively, knowing something about the degrees of each node can be used to sample graphs with the given degree sequences or expected degree distributions.

\paragraph{Noise \& missingness}
There is the potential for inaccuracies in the measurement of the cascade data.  In the above analysis we assume that there is no missing data; that is, all activations are observed and recorded correctly. 

There are three main sources of noise:
\begin{itemize}
\item Measurement errors: \eg missing or noisy meauresments - nodes that are activated are not recorded or recorded incorrectly.
\item Unobservable components: \eg outside influence: nodes are activated by information from an external source or node not in our network.
\item Model errors: \eg incubation time does not follow the assumed exponential distribution.
\end{itemize}

The Bayesian framework used here provides a natural way to deal with uncertainty and we can incorporate known missingness in the model. For instance, it is reasonable to assume noise in time measures, \eg $t = t_{\text{recorded}} + \epsilon$ with $\epsilon \sim \text{Exp}(\lambda)$. We can infer or make assumptions regarding the parameter $\lambda$, but we must take care if the errors switched the order of activation.

To incorporate unobservable nodes, many other works include a `source' node in the network that is connected to all nodes with some small probability $\gamma$ that provides external influence. This parameter could be constant over the network or depend on the node. Higher $gamma$ values increase the likelihood of external influence so reduce false negatives but increase false positives.  With some prior knowledge we can also estimate $\gamma$.

We have modelled the probability a node is activated from its neighbour as a simple exponential decay.  However, this can be extended to include other distributions as well as characteristics of the nodes and cascades.  For example, node degree as high degree nodes are potentially less likely to be influenced by a low degree node or some node similarity measure (are they `politically' or `socially' similar or produce similar textual content). Inferring some of these characteristics of nodes, such as their clustering, from information diffusion has recently been addressed in \cite{Peixoto-19} and remains an area of future work. The dependence of social influence and information flow on the content of the information could also be investigated and incorporated. There is also evidence for other distributions of waiting times, such as power-law with or without an exponential cutoff \cite{Gomez-12,Mathews-18}. These can easily be incorporated above.
%\change{CHECK IF IT HASNT ALREADY BEEN DONE.}
%\change{Not knowing the true parameters $\alpha$ We never actually did the power law incubation} 

\paragraph{Inferring $\beta_{uv}$ parameters that are dependent on the link itself}
There have been a number of works \cite{Embar-14,NETRATE,Peixoto-19} inferring the link strength between individuals as a measure of social influence or social trust using EM algorithms based on communication patterns.  More recently, Bayesian methods have been applied in this space. Embar \emph{et~al.} \cite{Embar-14} use a Bayesian framework to analytically and empirically infer link strengths in diffusion processes. The method here could be straightforwardly extended to include an inference of link strengths $\beta_{uv}$; however it is likely that much more data is required to infer strengths and strength distributions rather than simple adjacencies.

\paragraph{False positives and negatives}
The method reports marginal edge probabilities based on the cascades observed.  Highly probable edges that do not exist in $G^{*}$, or conversely edges of low probability that do, can still be informative. Non-edges with high probabilities are indicators that these edges provide alternate pathways through the network or may have been incorrectly measured in the ground truth.  Edges of low probability may be due to inactivity of the nodes or the presence of a highly probable edge that promotes the use of an alternate pathway. The presence or absence of edges in the posterior distribution tell us about the social influence pathways present in the network.  Even the incorrect classifications the algorithm makes can be informative when coupled with uncertainty quantification.

\paragraph{Time varying networks and data}
This algorithm assumes the underlying graph is static and all cascades have been observed.  It is widely understood that networks are time varying in many cases and often cascades are observed in an online fashion.  Online Bayesian inference methods like sequential Monte Carlo present an extension of this algorithm that can handle streaming data.  This stream could come in two forms: observing a cascade as it progresses over a network or observing a stream of full cascades when each concludes, or even both.  Additionally, online methods are capable of inferring dynamic network structures. As information is collected from these changing networks online algorithms will update posterior estimates of the changing network.  

\paragraph{Future work}
MCMC is known for slow convergence in high dimensional posteriors and can be memory intensive.  We have shown that despite this accurate and useful results can be obtained.  Other Bayesian inference methods are designed to combat these shortcomings. Sequential Monte Carlo, combined with importance sampling would not only allow online inference as discussed but also provide scope for more advanced proposals to improve acceptance rates.  Additionally, using an augmented model to allow deviations from the distribution of interest will improve mixing time and can be used in conjunction with importance sampling to return to the correct distribution.  Other cascade models allow for the likelihood to be parallelised over the nodes, rather than considering the graph as a whole. The approach can be easily extended to these models and MCMC chains run in parallel on each node will improve complexity.  These extensions provide exciting areas of future research.

\section{Conclusion}
We have presented an algorithm to infer the posterior distribution of networks that facilitate observed cascades.  Despite the high dimensional posterior basic MCMC methods can extract useful information and infer network structure based on samples from the distribution.  We also show that uncertainty quantification in networks, specifically when inferring networks from observed dynamics, has many benefits in not only the potential to improve informed decision making but also the ability to understand the complex interaction of network structures and the dynamics we observe.

\section{Acknowledgements}
The authors acknowledge the Data to Decisions CRC (D2D CRC), the Cooperative Research Centres Programme and the ARC Center of Excellence for Mathematical and Statistical Frontiers (ACEMS). This research is supported by an Australian Government Research Training Program (RTP) Scholarship.

\balance
%\section{References}
\bibliographystyle{acm}

\end{document}